\documentclass[twocolumn]{jpsj2} 
\usepackage{multicol}
\newcommand{\nc}{\newcommand}
\nc{\rnc}{\renewcommand}
\nc{\nn}{\nonumber}
\nc{\ch}{\cosh}
\nc{\sh}{\sinh}
\nc{\sech}{{\rm sech}}
\rnc{\Im}{{\rm{Im}\,}}
\rnc{\Re}{{\rm{Re}\,}}
\def\i{{\rm i}}
\def\e{{\rm e}}
\nc{\mfa}{{\mathfrak{a}}}
\nc{\mfab}{\overline{\mfa}}
\nc{\mfA}{{\mathfrak{A}}}
\nc{\mfAb}{\overline{\mfA}}
\nc{\mfb}{{\mathfrak{b}}}
\nc{\mfbb}{\overline{\mfb}}
\nc{\mfB}{{\mathfrak{B}}}
\nc{\mfBb}{\overline{\mfB}}
\nc{\mfsl}{{\mathfrak{sl}}}
\nc{\db}{\displaybreak[0]\\}
\nc{\bra}{\langle}
\nc{\ket}{\rangle}
\nc{\vf}{v_s}
\nc{\Dth}{D_{\rm Q}}
\nc{\Ds}{D_{\rm S}}
\nc{\tr}{{\rm Tr}}
\nc{\J}{\mathcal{J}}
\nc{\Jq}{\J_{\rm Q}}
\nc{\Je}{\J_{\rm E}}
\nc{\Js}{\J_{\rm S}}
\nc{\Jc}{\J_{\rm c}}
\nc{\lamr}{\Lambda_{\rm R}}
\nc{\laml}{\Lambda_{\rm L}}
\nc{\ep}{\varepsilon}
\rnc{\H}{\mathcal{H}}
\nc{\M}{\mathcal{M}}
\rnc{\(}{\left(}
\rnc{\)}{\right)}
\rnc{\d}{{\rm d}}
\nc{\lam}{\lambda}
\nc{\Lam}{\Lambda}
\nc{\DsI}{\Ds^{\rm I}}
\nc{\DsII}{\Ds^{\rm II}}
\nc{\dprime}{\prime\prime}

\numberwithin{equation}{section}

\title{Thermomagnetic Power and Figure of Merit 
for   Spin-1/2 Heisenberg  Chain}

\author{
Shunsuke \textsc{Furukawa},
Dai \textsc{Ikeda} and
Kazumitsu \textsc{Sakai}
}

\inst{Department of Physics, Tokyo Institute of Technology, 
            Tokyo 152-8551, Japan}
\abst{Transport properties in the presence of magnetic fields are 
numerically studied for the spin-1/2 Heisenberg XXZ chain. 
The breakdown of the spin-reversal symmetry due to the magnetic 
field induces the magnetothermal effect. 
In analogy with the thermoelectric effect in electron systems, the  
thermomagnetic power (magnetic Seebeck coefficient) is provided, 
and is numerically evaluated by  the exact diagonalization for wide 
ranges of temperatures and various magnetic fields.  For the 
antiferromagnetic regime, we find the magnetic Seebeck coefficient 
changes sign at certain temperatures,  which is interpreted as an 
effect of strong correlations.  We also compute the thermomagnetic 
figure  of merit determining the efficiency of the thermomagnetic 
devices for cooling or power generation.
}

\kword{transport properties, thermal conductivity, magnetothermal effect,
Seebeck coefficient, figure of merit, 
Heisenberg chain, quantum integrable system}

\begin{document}
\maketitle
%
\section{Introduction}
During the last two decades, strongly correlated  systems with 
reduced dimensions have been extensively studied from both 
theoretical and experimental perspectives, due to their unique  
static or dynamical properties.
Focusing our attention on the transport properties of magnetic 
materials described by one- or quasi one-dimensional  spin-1/2 
magnets, we find their unconventional features induced by 
magnetic excitations.
For instance, an unusually large spin diffusion constant has been 
observed in an NMR experiment for Sr$_2$CuO$_3$~\cite{Taki} 
which is well characterized by the  spin-1/2 Heisenberg chain.
In addition,  anomalously enhanced  thermal conductivities were 
also measured in experiments for Heisenberg chain compounds 
SrCuO$_2$ and Sr$_2$CuO$_3$,\cite{Sol1,Sol2} indicating a quasi-ballistic 
heat transport carried by the spinon or magnon with a mean-free path 
much larger than the correlation lengths.\cite{SODR}
Theoretically, these quasi-ballistic transport might be interpreted 
as a residual effect of integrability.
In fact the energy current  of a certain class of integrable systems 
including the spin-1/2 Heisenberg chain is written as a constant 
of motion, \cite{ZP03,znp} and therefore the heat transport  exhibits 
purely ballistic behavior.\cite{ZP03,znp,KS,SK} On the other hand 
the spin current is generally not conserved even in integrable systems, 
nevertheless it might have a finite overlap with the conserved quantities 
underlying integrability. Consequently the  spin current-current 
correlation does not decay to zero for long times, and then the 
spin transport is also considered to be 
ballistic.\cite{NarMA98,Zotos99,AlGros1,FK03} 
Although real materials such as Sr$_2$CuO$_3$ have non-integrable terms 
as a small perturbation, the system still exhibits the (quasi)-ballistic 
transport properties as already mentioned above. 
The integrability, however, is not a necessary condition to guarantee 
the existence of such ballistic transport properties. In fact, the 
quasi-ballistic heat transport was  also observed in spin-ladder 
compounds (Sr,Ca,La)$_{14}$Cu$_{24}$O$_{41}$,\cite{Kudo,Sol3,Kudo01,Hess01}
which can no longer  be explained by the integrability.
The arising question of what kind of non-integrable systems do 
or do not keep the (quasi)-ballistic features  still remains an 
intriguing open problem. \cite{AlGros1,FK03,MHCB,AlGros2,
OCC03,Saito1,Saito2,Saito3,MHCB2,SAR03,Zotos04,KZ04,Louis03}

Another crucial problem to be considered in the transport
properties is  effects of external fields.
By analogy with the thermoelectric effect in electron systems,
one might expect  the existence of the {\it magnetothermal}
effect in spin systems; the temperature gradient along a sample 
causes the magnetic field gradient.
For the one-dimensional spin systems without external fields, however,
there is no magnetothermal effect since the system exhibits 
the spin-reversal symmetry. 
In the presence of finite magnetic fields, the situation drastically 
changes; the magnetothermal effect indeed arises due to vanishing
of the spin-reversal symmetry.\cite{LG03,SK,MHCB3}
Thus for the complete understanding of the transport properties under 
finite external fields,  we must take into  account the magnetothermal 
effect correctly.
Quite recently, by utilizing the phenomenological approach, 
the {\it magnetic} Seebeck coefficient has been analytically calculated in 
the massless  regimes of the spin-1/2 Heisenberg 
XXZ chain.\cite{SK04}
Most significant, in that work, was the prediction that the
magnetic Seebeck coefficient changes sign at certain 
temperature and for certain  interaction strengths.
Unfortunately, the result is limited to the 
low-temperature region, due to the lack of conclusive results 
for the spin conductivity of the Heisenberg chain.

Motivated by this, in this paper, we will  discuss the 
magnetothermal effect for the spin-1/2 Heisenberg XXZ chain
beyond the limitation of the analytical approach.
Namely using the exact diagonalization method, we  evaluate 
the magnetic Seebeck coefficient for wide ranges of temperatures and 
various interaction strengths. As a consequence, we find the 
magnetic Seebeck coefficient changes sign for  interaction 
strengths in the antiferromagnetic regime, when the magnitude of the 
magnetic field satisfies a certain condition. Moreover we evaluate
the thermomagnetic figure of merit measuring the efficiency
of the thermomagnetic devices.

The layout of this paper is as follows. 
In the subsequent section, we briefly present a general formulation 
of the transport properties in spin systems. The transport coefficients 
and the thermomagnetic power are described within linear response 
theory. 
In section 3, we provide the spin and 
thermal currents for the spin-1/2 Heisenberg XXZ chain, and express 
the transport coefficients in terms of the correlation functions 
among the current operators. 
Numerical results by the exact diagonalization up to 18 sites are 
presented in section 4. Section 5 is devoted to the summary and
discussions.
In appendix, we shortly give  exact results for the free fermion 
(XY) model.
%
\section{Transport Coefficients in Spin Systems}
%
For later convenience, here we briefly provide a general 
formulation of the transport coefficients and the magnetothermal
effect in spin systems. 

Let us consider a system with two currents--the spin and  
heat currents--which flow as a result of  forces. 
The transport coefficients relate the currents to the 
driving forces, i.e. the potential and temperature 
gradients.
Phenomenologically these relations may be written in the form 
\cite{LG03,MHCB3}
\begin{equation}
\begin{pmatrix} \Js \\ \Jq
\end{pmatrix}
=
\begin{pmatrix}
L_{\rm SS} & L_{\rm SQ} \\
L_{\rm QS} & L_{\rm QQ}
\end{pmatrix}
\begin{pmatrix}
-\nabla \phi_s \\
-\nabla T/T
\end{pmatrix},
\label{response}
\end{equation}
where $\Js$, $\Jq$ are the spin and heat currents,
respectively, $\nabla \phi_s$ the potential
gradient (typically the magnetic field gradient
$\nabla \phi_s=-\nabla h$) and $\nabla T$ the temperature gradient.
The Kubo formula explicitly gives the coefficients 
$L_{ij}$ ($\{i,j\}\in\{\rm Q,S\}$) in terms of the correlation 
functions of the current operators: 
\begin{equation}
     L_{ij}=\lim_{\stackrel{\scriptstyle\omega\to0}{\epsilon\to+0}}
            \Re \int_0^{\infty}\d t\e^{-\i(\omega-\i\epsilon)t}
            \int_0^{\beta}\d\lambda
            \bra \J_i(-t-\i\lambda)\J_j \ket,
\label{kubo}
\end{equation}
where    $\{i,j\}\in\{\text{Q,S}\}$, $\beta$ is the reciprocal
temperature $\beta=1/T$ and $\bra \cdots \ket$ denotes
the thermal expectation value per site.
Using this relation  with eq.~\eqref{response},
one obtains the transport coefficients for the spin system.
The spin conductivity $\sigma$ is measured under the condition 
of no temperature gradient $\nabla T=0$:
\begin{equation}
\Js=\sigma (-\nabla \phi_s), \quad \sigma = L_{\rm SS}.
\label{spin-cond}
\end{equation}
On the other hand the thermal conductivity $\kappa$ is
defined when there is no spin current $\Js=0$:
\begin{equation}
\Jq=\kappa(-\nabla T), \quad
\kappa=\frac{1}{T}\left\{L_{\rm QQ}-\frac{ L_{\rm QS}^2}
{L_{\rm SS}}
\right\}.
\label{thermal-cond}
\end{equation}
Note that here we have imposed the Onsager relation $L_{\rm SQ}
=L_{\rm QS}$. 

On the analogy of the thermoelectric power (Seebeck coefficient) 
for electron systems, we define the thermomagnetic power 
(we refer to as the ``magnetic" Seebeck coefficient) $S$, 
which should be measured under the condition $\Js=0$:
\begin{equation}
S=-\frac{\nabla \phi_s}{\nabla T}=
\frac{1}{T}\frac{L_{\rm QS}}{ L_{\rm SS}}.
\label{Seebeck1}
\end{equation}
The (magnetic) Seebeck coefficient is  a  crucial 
criterion to clarify the types of carriers. Namely when the 
sign of $S$ is positive (negative), the spin and heat
are dominantly carried by the carriers with up-spin (down-spin).

Finally as another important quantity  which  
determines the efficiency of thermomagnetic devices for cooling 
or power generation, we define the thermomagnetic figure of merit
\begin{equation}
Z T=\frac{S^2 \sigma }{\kappa } T=
\frac{L_{\rm Q S}^2 }{L_{\rm QQ}L_{\rm ss}-L_{QS}^2}.
\label{FOM1}
\end{equation}

%
\section{Current Operators in the Heisenberg chain}
\begin{figure}[hh]
\begin{center}
\includegraphics[width=0.45\textwidth]{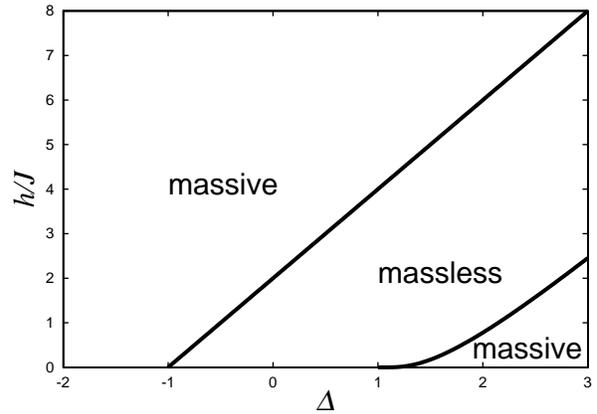}
\end{center}
\caption{Phase diagram of the ground state for $h>0$ and
$J>0$.}
\label{PD}
\end{figure}

%
Here we apply the formulae given in the preceding section
to the spin-1/2 Heisenberg XXZ chain.
The Hamiltonian  is defined as
\begin{align}
   &\H=\sum_{k=1}^{L}h_{kk+1}-\frac{h}{2}\sum_{k=1}^{L}\sigma_k^z,\nn \\
    &h_{kk+1}=J\left\{\sigma_{k}^+\sigma_{k+1}^-
                     +\sigma_{k+1}^+\sigma_{k}^-
                     +\frac{\Delta}{2}(\sigma_k^z\sigma_{k+1}^z-1)
              \right\},
\label{hamiltonian}
\end{align}
where $\sigma_k^{\pm}=(\sigma_k^x\pm\i\sigma_k^y)/2$ and 
$\sigma_k^{a}$ ($a\in\{x,y,z\}$) are the Pauli matrices
associated with the $k$th site of the chain.
{}From now on,  we assume $L$ is even, and impose the periodic boundary 
condition ($\sigma^{a}_{1}=\sigma^{a}_{L+1}$). The coupling 
constants $J$ and $\Delta$ together with the magnetic field $h$ 
determine the ground state properties of the system \cite{Takabook} 
(see Fig.~\ref{PD}).
Since the energy spectrum is invariant under both the
transformations $(h,\M=\sum_k\sigma_k^z/2)\leftrightarrow(-h,-\M)$ and 
$(J,\Delta)\leftrightarrow(-J,-\Delta)$, we assume $h>0$ and $J>0$.

To evaluate eq.~\eqref{kubo}, we define the spin and energy 
currents from the continuity equations for the local 
spin $S_k^z=\sigma_k^z/2$ and energy $h_{k k+1}$:\cite{ZP03}
\begin{equation}
\dot{S}_k^z=-{\rm div} j_k^{\rm S},\quad
\dot{h}_{k k+1}=-{\rm div} j^{\rm E}_k,
\end{equation}
where ${\rm div} j_k:=j_{k+1}-j_k$. These relations give the 
spin and energy current, $\Js=\sum_k j_k^{\rm S}=
\sum_k \i[h_{k-1k},S_k^z]$ and
$\Je=\sum_k j_k^{\rm E}=
\sum_k \i [h_{k-1k},h_{kk+1}] $, respectively. Their explicit
forms read
\begin{align}
    \Js&=\i J\sum_{k=1}^{L}(\sigma_k^+\sigma_{k+1}^-
          -\sigma_{k+1}^+\sigma_{k}^-), \nn \\
    \Je&=-\i J^2  \sum_{k=1}^{L}  
           \biggl\{\sigma_k^{z}(\sigma_{k-1}^{+}\sigma_{k+1}^{-}-
                   \sigma_{k+1}^{+}\sigma_{k-1}^{-}) \nn \\
      & \quad -\Delta(\sigma_{k-1}^{z}+\sigma_{k+2}^{z})
                 (\sigma_{k}^{+}\sigma_{k+1}^{-}-
                  \sigma_{k+1}^{+}\sigma_{k}^{-})\biggr\}.
\end{align}
Note that the  heat current $\Jq$ should be 
defined as
\begin{align}
\Jq=\i\sum[h_{k-1,k}-h S^z_{k-1},
h_{k,k+1}-h S^z_{k}]=\Je-h\Js.
\label{heat}
\end{align}

In general, the exact evaluation of the dynamical correlation
functions such as eq.~\eqref{kubo} is highly challenging
problem even in exactly solvable models. 
However, the energy current of a certain class of solvable 
models including the present system is written as one of 
the non-trivial conserved quantities underlying integrability, 
\cite{znp} i.e.  $[\H,\Je]=0$. This important fact directly 
leads to the diverging thermal conductivity at zero frequency
$\omega=0$:
\begin{equation}
   \kappa=\pi \Dth \delta(\omega),
\label{thermal-cond2}
\end{equation}
where the weight of the delta function $\Dth$ is 
referred to as the thermal Drude weight\cite{KS,SK,LG03,MHCB3,SK04}
\begin{equation}
\Dth=\beta^2 \bra \Je ^2\ket-
                 \beta^3 \frac{\bra \Je \Js \ket^2}{\Ds}.
   \label{thermalDrude}
\end{equation}
Here $\Ds$ is the Drude weight for the spin transport, 
\begin{equation}
\sigma=\pi \Ds\delta(\omega)+\sigma ^{\rm reg}.
\label{spinDrude}
\end{equation}

For zero magnetic field $h=0$, the magnetothermal effect,
which is related to the off-diagonal dynamical correlation 
$L_{\rm QS}(=L_{\rm SQ})$, is always zero 
(i.e. $\bra\Je \Js \ket=0$) because the system exhibits the 
spin-reversal symmetry. In this case the thermal Drude weight
is simply written as $\Dth=\beta^2 \bra \Je^2 \ket$, and 
can be exactly  evaluated by  the Bethe ansatz technique.
\cite{KS,SK}

In contrast, for finite magnetic fields $h>0$, the magnetothermal 
effect arises due to vanishing of the spin-reversal symmetry,
$\bra \Je\Js \ket>0$.  
In addition to this, the spin conductivity also diverges
(i.e. $\Ds>0$) as far as $h>0$, which is  proven
by using the Mazur inequality \cite{znp}
\begin{equation}
\Ds\ge  \frac{\bra \Je \Js \ket^2}{T \bra \Je^2\ket}>0 \quad
\text{for $h>0$}.
\end{equation}
Hence the thermal Drude weight \eqref{thermalDrude} and
the magnetic Seebeck coefficient (cf. \eqref{Seebeck1}) 
\begin{equation}
   S=\frac{1}{T}\left\{\frac{\bra \Je\Js
   \ket}{\Ds}\frac{1}{T}-h\right\},
\label{Seebeck2}
\end{equation}
are both finite at finite temperatures and magnetic fields. 
Insertion of eqs.~\eqref{thermal-cond2}, \eqref{spinDrude} 
into eq.~\eqref{FOM1} yields
\begin{equation}
ZT=\frac{S^2 \Ds }{\Dth} T.
\label{FOM2}
\end{equation}

To analyze the magnetothermal effect for the XXZ chain 
\eqref{hamiltonian}, we must explicitly  determine the spin 
Drude weight as well as the correlations $\bra\Je^2 \ket$ and 
$\bra\Je\Js \ket$.
In fact, both the quantities $\bra \Je\Js \ket$ and $\bra \Je^2 \ket$ 
have already evaluated by the Bethe ansatz in ref.~\citen{SK04}.
On the other hand, for the spin Drude weight, we may derive $\Ds$ 
by considering the finite size corrections of the  ``string" 
solutions to the Bethe ansatz equation.\cite{FK98,Zotos99}
In general, however, the validity of the results obtained by 
applying the finite size correction to the string solutions 
are highly questionable under the circumstance that the breakdown 
of the string hypothesis is reported as 
in ref.~\citen{EKSstring,JuDo,AlczMart}. 
Indeed the resultant 
Drude weight $\Ds$ derived in ref.~\citen{Zotos99} for the case of
zero magnetic field is inconsistent 
with that obtained by numerical analysis \cite{AlGros1} 
or field theoretical arguments.\cite{FK03}
Consequently the exact analysis of $\Ds$ at arbitrary temperatures
still leaves an open problem.

Alternatively, in the next section,  we will calculate the
magnetic Seebeck coefficient $S$ and the figure of merit 
$ZT$ by the exact diagonalization for finite chains.

Finally we comment on the transport properties of the
spinless fermion system given by
performing the Jordan-Wigner transformation on the 
XXZ  chain \eqref{hamiltonian}:
\begin{align}
\H_{\rm c}&=\sum_{k=1}^L\biggl\{ t(c_k^{\dagger}c_{k+1}+c_{k+1}^{\dagger}c_k)
+V \bigl(n_k-\frac{1}{2}\bigr) \bigl(n_{k+1}-\frac{1}{2}\bigr)\biggr\} \nn \\
&\qquad -\mu \sum_{k=1}^L\bigl(n_k-\frac{1}{2} \bigr),
\label{hamiltonian2}
\end{align}
where $c_k^{\dagger}$ ($c_k$) is the fermionic creation (annihilation)
operator on the site $k$ and $n_k=c_{k}^{\dagger} c_k$. Note
that we have also transformed  $J \to t$, $h\to\mu$ and 
$\Delta \to V/(2t)$. The corresponding charge 
and energy currents are respectively given by
\begin{align}
    \Jc&=e t\sum_{k=1}^{L}(\i c_k^\dagger c_{k+1}+{\rm h.c.}), \nn \\
    \Je&=  t \sum_{k=1}^{L}  
           \biggl\{t (\i c_{k-1}^{\dagger}c_{k+1}+{\rm h.c.}) \nn \\
 & \quad +V (n_{k+1}+n_{k+2}-1)
              (\i c_k^\dagger c_{k+1}+{\rm h.c.})\biggr\},
\end{align}
where $e$ is the charge of the particle. Accordingly the
heat current is defined as $\Jq=\Je-(\mu/e) \Jc$ (cf. \eqref{heat}). 
In the thermodynamic
limit $L\to\infty$ where the difference of  boundary conditions
between the spin system and the corresponding fermion system vanishes,
the charge (thermal) conductivity $\sigma_{\rm c}$ ($\kappa_{\rm c}$), 
the Seebeck coefficient $S_{\rm c}$
and the thermoelectric figure of merit $Z_{\rm c} T$ are respectively 
related to those of 
the spin system:
\begin{equation}
\sigma_{\rm c}=e^2\sigma, \quad \kappa_{\rm c}=\kappa, \quad 
S_{\rm c}=\frac{S}{e}, \quad Z_{\rm c}T= Z T.
\label{fermion}
\end{equation}

%
\section{Numerical Analysis for magnetic Seebeck Coefficients and
Figures of Merit}
%
In this section we evaluate the magnetic Seebeck coefficient $S$
and the thermomagnetic figure of merit $ZT$ by the exact
diagonalization up to 18 sites.

Before we analyze them in detail, let us shortly review 
the magnetothermal effect for several special cases, i.e. 
the XY ($\Delta=0$) limit, and the low-temperature 
limits, where the analytical solutions are available.

\subsection{XY model}
For the XY model, both the spin and thermal currents are conserved and
hence the magnetothermal effect is exactly calculated 
by using the Jordan--Wigner transformation (see Appendix).
In Fig.~\ref{XY}, the magnetic Seebeck coefficient $S$ together
with the thermomagnetic figure of merit $ZT$ are depicted
for various magnetic fields.
In this case, one finds the magnetic Seebeck coefficients are
always negative ($S<0$) at any finite temperatures, which 
implies  the carriers of the transport are dominated
by the down spins.
%
%
\begin{figure}[hh]
\begin{center}
\includegraphics[width=0.45\textwidth]{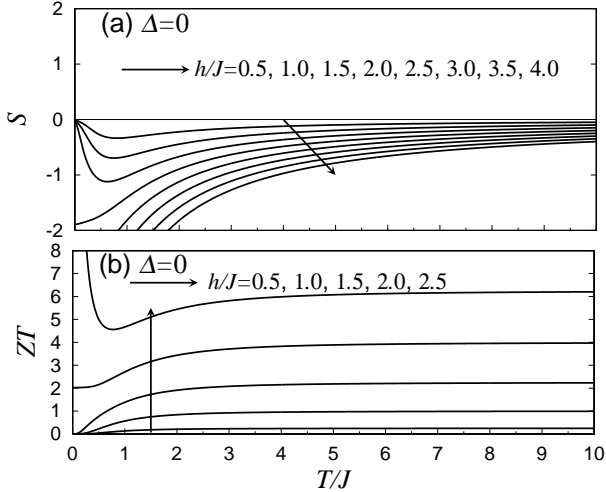}
\end{center}
\caption{Magnetic Seebeck coefficient $S$ (a) and 
figure of merit $ZT$ (b) for the XY model $\Delta=0$.
For $h=h_c=2J$, $S$ and $ZT$ converge to $S=-1.89738$
and $ZT=2.02697$, respectively (see eq.~\eqref{LT-fermi}).
}
\label{XY}
\end{figure}
%

The magnetic Seebeck coefficient $S$ for $h<h_c=2J$
is linear in $T$ at low temperature.
One also observes that $S$
diverges at $T=0$ as $S\sim {\rm const.}/T$ (see eq.~\eqref{LT-fermi}
in detail) for $h>h_c$, due to the mass-gap 
(spin insulator).
Correspondingly the thermomagnetic figure of merit $ZT$
increases with increasing the magnetic fields and diverges
as $ZT\sim {\rm const.}/T^2$ at $T=0$ for $h>h_c$.
\subsection{Low-temperature asymptotics for $-1<\Delta\le 1$}
Next we mention the low-temperature behavior $T\ll h <h_c$
in the presence of the interaction strengths $-1<\Delta\le 1$.
The phenomenological relation yields
the leading low-temperature magnetic Seebeck coefficient 
for $T\ll h<h_c$ \cite{SK04}:
\[
S=\alpha(h) T+O(T^2).
\]
%
\begin{figure}[hh]
\begin{center}
\includegraphics[width=0.45\textwidth]{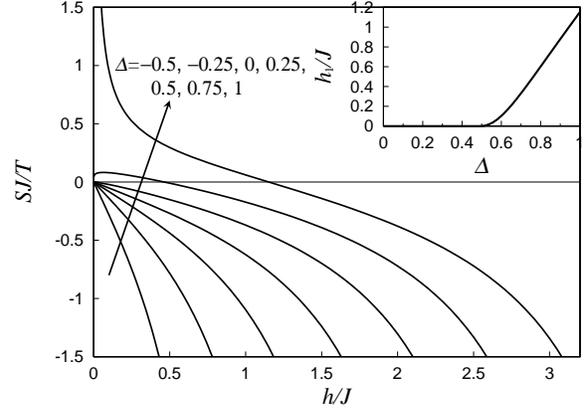}
\end{center}
\caption{Low-temperature behavior of the magnetic Seebeck
coefficient $S$ in units of $T/J$ for critical 
regime $-1<\Delta\le 1$. Inset: The magnetic field 
$h_1$ where the low-temperature asymptotics changes sign
is shown as the function of $\Delta$.}
\label{LT-Seebeck}
\end{figure}
%
In Fig.~\ref{LT-Seebeck}, the coefficient  $\alpha(h)$  is 
depicted as a function of the magnetic field for various anisotropy 
parameters. For weak interaction strengths $\Delta\lesssim 0.5$, 
the leading behavior is negative. On the contrary, for $\Delta\gtrsim0.5$, 
the low-temperature behavior changes sign at certain magnetic field 
$h=h_1$ (see the inset in Fig.~\ref{LT-Seebeck}). 
The value $h_1$ shifts to higher values with the increase of  the 
interaction strength. 
On the other hand, the magnetic Seebeck coefficient at
 high temperatures is determined according to the 
argument as in ref.~\citen{CB}:
\begin{equation}
S=-\frac{h}{T} \qquad \text{for $T\gg J$}.
\label{hT}
\end{equation}
{}From this relation and the results in Fig.~\ref{LT-Seebeck},
we expect $S$  has a positive peak and changes sign at
certain temperature $T_0$ at least for $\Delta\gtrsim 0.5$ and
$h<h_1$. Namely the crossover from the regime dominated by the 
``down-spin"-like carriers to the regime dominated by 
the ``up-spin"-like carriers  occurs at $T_0$.

In the next subsection,  beyond the limitation of 
the analytic methods, we numerically calculate the 
temperature dependence of the magnetic Seebeck coefficients
for various magnetic fields and anisotropies. 
%
\subsection{Exact diagonalization}
%
We next present our numerical results obtained by 
the exact diagonalization (ED) method.  
Using the conservation of the total magnetization 
$\mathcal{M}$ and the translational invariance of
the system, we performed a full diagonalization  up to $L=18$.  
%
To this end, we reduced the integral representation 
of the spin Drude weight $\Ds$ in \eqref{spinDrude}
(see also \eqref{spin-cond} and \eqref{kubo}) to a useful form.
The spectral decomposition $\H |m\ket =E_m |m \ket$
directly yields
\begin{equation}
\DsI=\beta \sum_{E_m=E_n}p_m |\bra m|\Js |n\ket|^2, \quad
p_m=\frac{1}{L}\frac{\e^{-\beta E_m}}{\sum_n \e^{-\beta E_n}}, 
\label{drude1}
\end{equation}
Applying the partial integration to $L_{\rm SS}$ \eqref{kubo}, 
one also derives 
\begin{align}
\DsII=\bra -K \ket -2\sum_{E_m\ne E_n}p_n \frac{|\bra m| 
\Js |n \ket|^2}{E_m-E_n},
\label{drude2}
\end{align}
where $K$ is the kinetic term 
$K=J \sum_k (\sigma_k^+\sigma_{k+1}^{-}+\text{h.c.})$. 
Here we have used the relation
\begin{equation}
\int_0^{\beta}\bra \Js(-\i \lam)\Js \ket\d \lam=\bra -K \ket,
\end{equation}
which is valid in the thermodynamic limit $L\to\infty$. 
%
\begin{figure}[hh]
\begin{center}
\includegraphics[width=0.45\textwidth]{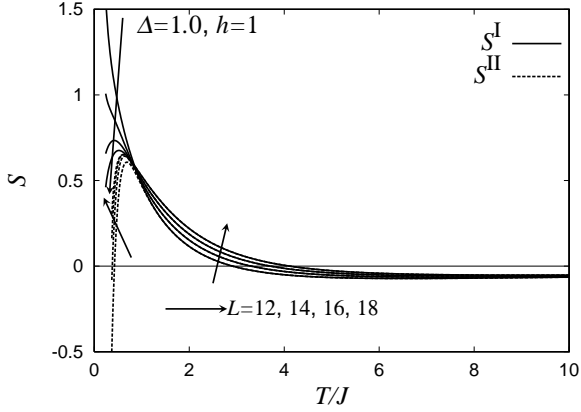}
\end{center}
\caption{
Size dependence of the magnetic Seebeck coefficient $S$ for 
$\Delta=1$ and $h/J=1$}
\label{fse}
\end{figure}
%
In Fig.~\ref{fse}, the size dependence of the magnetic
Seebeck 
coefficient $S^{\rm I}$ ($S^{\rm II}$) 
calculated by using  $\DsI$ ($\DsII$). 
$S^{\rm I}$ and $S^{\rm II}$ show a good agreement at high
temperatures, confirming $\DsI$ \eqref{drude1} and $\DsII$ 
\eqref{drude2} are equivalent at high temperature limit. 
In this region, 
one also observes a relatively weak size dependence.
On the other hand, at low temperatures, both $S^{\rm I}$
and $S^{\rm II}$ exhibit  strong size dependences,
which mainly stem from the strong finite size correction
of the spin Drude weight. In this region, one sees
$S^{\rm I}$ ($S^{\rm II}$) decreases (increases) with 
increasing the system size. {}From this observation, 
we  expect the magnetic Seebeck coefficient $S$ in the 
thermodynamic  limit converges to an intermediate 
value of $S^{\rm I}$ and $S^{\rm II}$.

\begin{figure}[ttt]
\begin{center}
\includegraphics[width=0.45\textwidth]{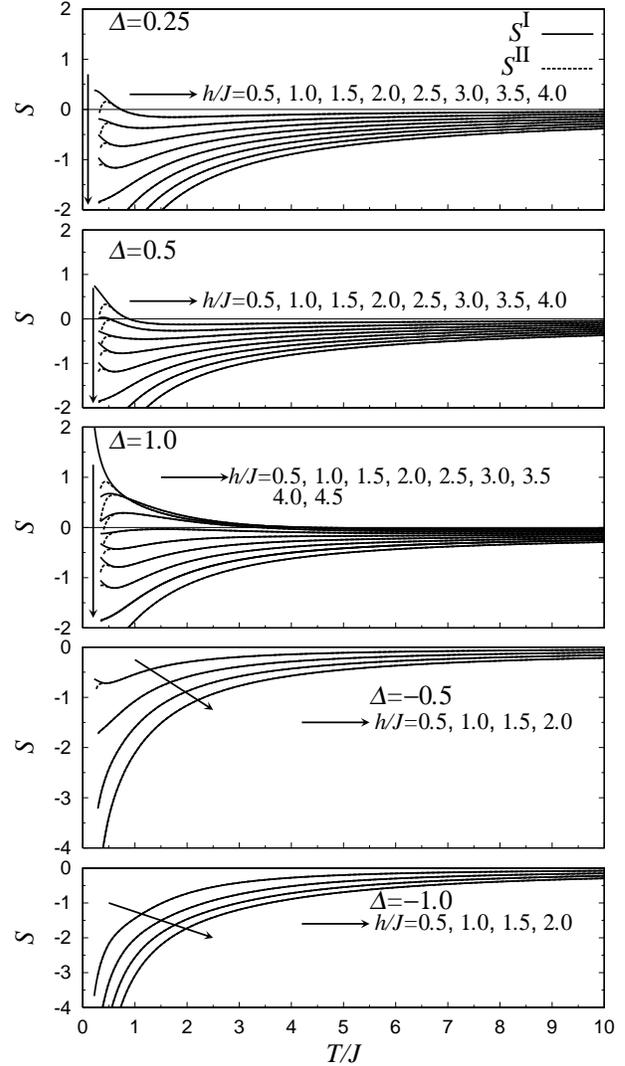}
\end{center}
\caption{Temperature dependence of the magnetic Seebeck coefficients
$S$ for both $\Delta>0$ and $\Delta<0$.}
\label{S}
\end{figure}

The magnetic Seebeck coefficient $S$ for several anisotropies 
$\Delta$ and magnetic fields  $h$ are shown in Fig.~\ref{S}.  
One observes that $S$ strongly depends on $h$ and 
$\Delta$. 
As expected in eq.~\eqref{hT}, $S$ is negative at high 
temperatures, and converges to zero at $T\to\infty$.

Above the critical field $h>h_c$ where all the spins point up 
at $T=0$, the behavior of $S$ is 
similar to that of the XY model. Namely $S$ monotonously 
decreases with increasing $h$. At low temperature regime, 
$S$ converges to a certain finite value for $h=h_c$, and 
diverges like $S\sim -\delta/T$ for $h>h_c$, where $\delta$ 
is the one-magnon excitation gap (see later discussion).

%
%

For $h<h_c$, in contrast, the behavior of the magnetic Seebeck coefficients
in the regime $\Delta>0$ is clearly different from that for 
$\Delta<0$. Namely for $\Delta>0$, $S$  has a positive
peak (in other words $S$ changes sign) below certain magnetic 
field $h<h_1$, while for $\Delta<0$, $S$ is always negative 
as in the XY model.
This behavior is consistent with the former 
prediction mentioned in the preceding subsection. 
In our investigation, however, we find that this sign changes 
occur even in $0<\Delta \le 0.5$, which cannot be 
concluded as long as we concentrate on the low-temperature 
behavior since the low-temperature asymptotics is negative
in this region (cf. Fig.~\ref{LT-Seebeck}).
One can also find in Fig.~\ref{S} for $\Delta>0$ 
that the positive 
peak grows in height, and the temperature range where 
$S>0$ becomes wide with the increase of $\Delta$.

\begin{figure}[hh]
\begin{center}
\includegraphics[width=0.45\textwidth]{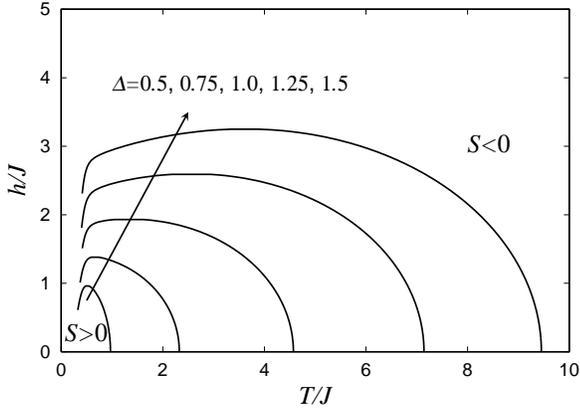}
\end{center}
\caption{Boundary where the magnetic Seebeck coefficient $S$ 
changes sign.}
\label{diagram}
\end{figure}

In Fig.~\ref{diagram} the boundary on which $S$ changes sign 
is shown as a line on the $h-T$ plane. The inside (outside)
of the curve denotes the region $S>0$ ($S<0$).
For any $\Delta>0$, the line has a maximum at a certain 
finite temperature. This feature together
with the low-temperature asymptotics shown in 
Fig.~\ref{LT-Seebeck} indicate that, for $h_1<h<h_2$, 
the sign change occurs at the temperatures $T_1$ and 
$T_2$ ($S(T)>0$ for $T_1<T<T_2$ 
and $S(T)<0$ for $0<T<T_1$ or $T>T_2$), 
while for $0<h<h_1$, it occurs at only one point 
$T=T_0$ ($S(T)>0$ for $T<T_0$ and $S(T)<0$ for $T>T_0$).
Here $h_2$  is the height of the maximum of the line
depicted in Fig.~\ref{diagram} and $h_1$ is defined in the preceding 
section (see Fig.~\ref{LT-Seebeck}).

%
%
This remarkable feature may be qualitatively interpreted
by considering the following extreme cases.
(i) For $h\gg 1$ where the almost every spin points up,
the elementary excitation is described by the
down-spin magnon. In this case, the mobility of the 
down-spin magnon may be larger than that of the 
up-spin magnon, implying the negative magnetic Seebeck coefficient 
$S<0$.
(ii) For $\Delta\gg 1$ and $0<h\ll 1$, the excitation from 
the ground state is mainly characterized by the up 
spinon, which also plays a role as a carrier, 
indicating the positive magnetic Seebeck coefficient, i.e. $S>0$. 
(iii) For $T\gg 1$ where the
interaction $\Delta$ is irrelevant, the behavior of $S$
is similar to the XY case where $S<0$.

For the realistic regimes $\Delta, T, h \sim 1$, the
sign changes of $S$ might be interpreted as the competition
of the above three mechanisms. Namely the sign change
at $T=T_0$ or $T=T_2$ is explained in terms of
(ii) and (iii). On the other hand, the sign change at
$T=T_1$ is described by (i) and
(ii), i.e. the crossover from the regime dominated by
the  carriers of down magnons  to the regime dominated
by the carriers of up spinons  occurs at $T=T_1$.

\begin{figure}[hh]
\begin{center}
\includegraphics[width=0.45\textwidth]{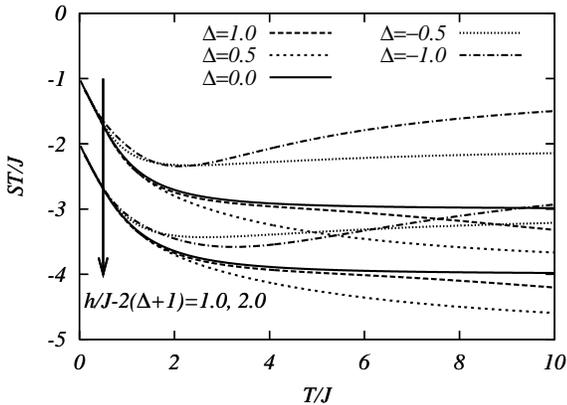}
\end{center}
\caption{The magnetic Seebeck coefficient $S$ multiplied
by $T/J$ above the critical fields. In the low-temperature
limit, $S$ behaves as $S\sim -\delta/T$ where $\delta$ is
the one-magnon excitation gap $\delta=h-2J(\Delta+1)$.}
\label{mass}
\end{figure}

We next investigate in detail the divergent behavior 
of $S$ above the saturation field $h_c$.  As shown in 
Fig.~\ref{mass}, the divergence of $S$ at low temperature 
can be canceled by multiplying it by $T/J$.  The value of $ST/J$ 
approaches  $-\delta/J$ in the low-temperature limit, 
where $\delta$ is the one-magnon excitation gap 
$\delta=h-2J(\Delta+1)$.  Thus we propose that the leading 
asymptotics at low temperature is written as $S=-\delta/T$.  
%

\begin{figure}[hh]
\begin{center}
\includegraphics[width=0.45\textwidth]{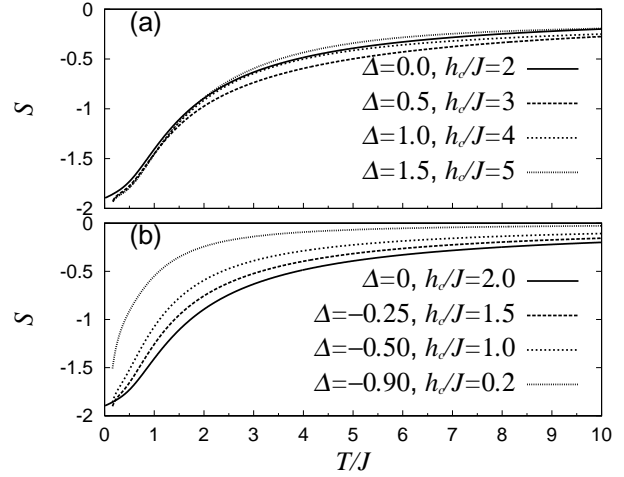}
\end{center}
\caption{Temperature dependence of the magnetic Seebeck
coefficient $S$ for the critical field $h_c=2J(1+\Delta)$}
\label{critical}
\end{figure}

For $h=h_c$ where $S$ converges to a finite value 
in the low-temperature limit, we observe a universal 
temperature dependence of $S$ for $\Delta>0$ 
as shown in Fig.~\ref{critical}(a), i.e. the behavior
of $S$ is well described by that
of the non-interacting case.
The existence of such a universal behavior at $h=h_c$ has also 
be pointed out by Heidrich-Meisner {\it et al.} 
in the investigation of the thermal Drude weight \cite{MHCB3}.  
For $\Delta<0$ in Fig.~\ref{critical}(b), in contrast, the temperature 
dependence of $S$ strongly depends on the anisotropy.
%
%
%
%
\begin{figure}[hh]
\begin{center}
\includegraphics[width=0.45\textwidth]{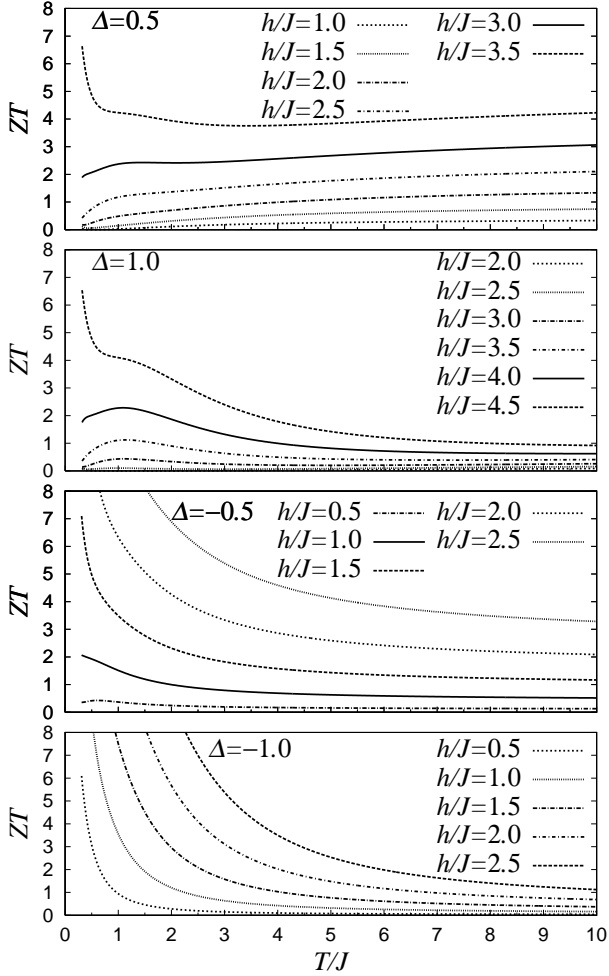}
\end{center}
\caption{Temperature dependence of the figure of merit for 
$ZT$ both $\Delta>0$ and $\Delta<0$.}
\label{FOM}
\end{figure}
%

Finally we discuss  our numerical results for the
thermomagnetic figure of merit $ZT$. 
The temperature dependence of $ZT$ obtained by 
using $\DsII$ is depicted in  Fig.~\ref{FOM} 
for various magnetic fields and several 
anisotropies. 
For $h<h_c$, the figure of merit $ZT$ approaches zero
at zero temperature. On the other hand,  
$ZT$ converges to a certain finite value
for  $h=h_c$, and diverges for $h>h_c$,
which is similarly observed in the XY case. 
In general, for fixed magnetic field, 
$ZT$ in $\Delta>0$ tends to decrease with the 
increase of the interaction strength $\Delta$, reflecting
the fact that the thermal Drude weight 
increases, while the spin Drude weight decreases 
with increasing $\Delta$.
%
\section{Summary and Discussion}
%
In this paper, the magnetothermal effect was studied for 
both the ferromagnetic and antiferromagnetic regimes of the XXZ 
chain with external magnetic fields. The magnetic Seebeck coefficients 
and the figures of merit  were numerically calculated by using the exact 
diagonalization up to 18 sites. For the 
antiferromagnetic regime, we found the magnetic Seebeck coefficient 
changes sign at certain temperatures,  which is interpreted as an
effect of the competition among the interaction strengths, temperatures
and external magnetic fields.

An extension to  integrable correlated electron 
systems with the spin degree of freedom 
such as the Hubbard model and the supersymmetric $t-J$ model
is an interesting problem. Due to the integrability, the
thermal and the spin Drude weights are considered to be finite 
at finite temperatures. In particular, the energy current
for the supersymmetric $t-J$ model is conserved and hence
our method developed in this paper can be directly
applied. For zero magnetic fields, Zemljic and Prelov\v sek
have recently investigated the thermoelectric power of the
Hubbard model.\cite{MP04} Because of   strong correlations,
the sign changes of the (electric) Seebeck coefficient have been observed 
as in the present case.
 In the presence of external magnetic fields, 
the magnetothermal effect arises and crucially affects the 
charge transport and the thermoelectric effect. The investigation 
of this effect is also of importance for better understanding
of the transport properties of strongly correlated 
electron systems with reduced dimensions.

\section*{Acknowledgment}
We would like to thank A. Kl\"umper and
M. Oshikawa for stimulating discussions.
This work is supported in part by a 21st Century COE Program at
Tokyo Tech ``Nanometer-Scale Quantum Physics'' by the
Ministry of Education, Culture, Sports, Science and Technology.
KS is also supported by Grant-in-Aid for Young Scientists (B)
No. 17740248.

%
%
\appendix
\section{XY  Case}
%
In this appendix, we provide the exact results of 
the transport coefficients and the thermomagnetic power
for the XY chain, which corresponds
to $\Delta=0$ in eq.~\eqref{hamiltonian}. 
In this case, the spin current $\Js$ is a constant of
motion, i.e. $[\H,\J_s]=0$. Hence from eq.~\eqref{drude1},
one sees that the spin Drude weight is simply expressed
as $\Ds=\beta \bra \Js^2\ket$.
To calculate the correlation functions 
$\bra\Je^2\ket$ and $\bra\Je\Js\ket$,
we introduce the partition function $Z$ as
\begin{equation}
\ln Z(\lam_1,\lam_2)=
\frac{ \ln \tr \exp\{-\beta \H+\lam_1 \Js+\lam_2\Je\}}{L}.
\label{partition}
\end{equation}
Taking the derivative of this with respect to 
the parameters $\lam_1$ and $\lam_2$, and 
then setting $(\lam_1,\lam_2)=(0,0)$, one obtains the 
desired quantities:
\begin{align}
&\Ds=\beta \bra \Js^2 \ket=
\partial^2_{\lam_1} \ln Z(\lam_1,0)\bigr|_{\lam_1=0}, \nn \\
&\bra \Je^2 \ket =\partial_{\lam_2}^2 \ln Z(0,\lam_2)
\bigr|_{\lam_2=0},\quad \nn \\
&\bra \Je \Js \ket =\partial_{\lam_2}\partial_{\lam_1} 
\ln Z(\lam_1,\lam_2)\bigr|_{\lam_1=0,\lam_2=0}.
\label{currentXY}
\end{align}
For actual evaluation of $Z(\lam_1,\lam_2)$, it is
convenient to transform the system to the spinless 
fermion model \eqref{hamiltonian2} (note that we set $e=1$).
Performing the Fourier transform, one easily obtains
\begin{align}
\H&=2 J \sum_p n_p \cos p -h \sum_p \bigl(n_p-\frac{1}{2}\bigr), \nn \\
\Js&=-2  J\sum_p n_p \sin p,  \,\
\Je=-2 J^2 \sum_p n_p \sin 2p,
\end{align}
where $p=2\pi  n/ L$; $n\in\{-L/2+1,-L/2,\cdots, L/2-1,L/2\}$.
Substituting them into \eqref{partition} and taking the
thermodynamic limit $L\to\infty$, we obtain
\begin{align}
&\ln Z(\lam_1,\lam_2)=-\frac{\beta h}{2} +\frac{1}{2\pi}
\int_{-\pi}^{\pi}\ln(1+\e^{\varepsilon(p)})\d p, \nn \\
&\varepsilon(p)=-2\beta J \cos p-2\lam_1 J \sin p-2\lam_2 J^2 \sin 2p+\beta h.
\end{align}
Thus eq.~\eqref{currentXY} yields
\begin{align}
&\Ds= \frac{1}{2\pi}\int_{-\pi}^{\pi}
   \frac{ \beta J^2   \sin^2 p}
        { \cosh^2( \beta J \cos p- \beta h/2)}\d p,\nn \\
&\bra \Je^2 \ket= \frac{1}{2\pi}\int_{-\pi}^{\pi}
   \frac{J^4 \sin^2 2p}
          { \cosh^2( \beta J \cos p- \beta h/2)}\d p,\nn \\
&\bra \Je\Js \ket=\frac{1}{2\pi}\int_{-\pi}^{\pi}
   \frac{J^3 \sin p \sin 2p}
        { \cosh^2( \beta J \cos p- \beta h/2)}\d p. \nn 
\end{align}
Combining the above equations with 
\eqref{thermalDrude}, \eqref{Seebeck2} and \eqref{FOM2},
one can calculate the thermal Drude weight and the magnetic
Seebeck coefficient for arbitrary temperatures and magnetic fields.
The low-temperature asymptotics of the magnetic Seebeck
coefficient and the figure of merit are explicitly written in
the following form.
\begin{align}
S&=\begin{cases}
          -\dfrac{\pi^2 h }{3 v_s^2 \beta}+O(\beta^{-3})
                            &\,\, \text{for $h<h_{c}$} \\[4mm]
          -\dfrac{g(3)}{g(1)}+O(\beta^{-1})
                             &\,\, \text{for $h=h_c$} \\
          (2J-h)\beta+O(1)
                             &\,\, \text{for $h>h_c$}
    \end{cases}, \nn \\
\quad
ZT&=\begin{cases}
          \dfrac{\pi^2 h }{3 v_s^2 \beta^2}+O(\beta^{-4})
                            &\,\, \text{for $h<h_{c}$} \\[4mm]
          \dfrac{g^2(3)}{g(1)g(5)-g^2(3)}+O(\beta^{-1})
                            & \,\, \text{for $h=h_c$} \\[4mm]
          \dfrac{2}{3}(2J-h)^2\beta^2+O(\beta)
                            & \,\, \text{for $h>h_c$} 
    \end{cases},
\label{LT-fermi}
\end{align}
where $\vf=\sqrt{4J^2-h^2}$ is the velocity of excitations,
$h_c=2 J$ and
$g(n):=(1-2^{1-n/2})\varGamma(1+n/2)\zeta(n/2)$
(note that $\varGamma(x)$ is the Gamma function and $\zeta(x)$  
the Riemann  zeta function).
On the other hand, the high-temperature asymptotics are given by
\begin{align}
S\sim-h \beta + O(\beta^2),\quad
ZT\sim\frac{h^2}{J^2}.
\end{align}
%


\begin{thebibliography}{99}
%
\bibitem{Taki} M.~Takigawa,{\it et al}.:
Phys. Rev. Lett. {\bf 76} (1996) 4612.
%
\bibitem{Sol1}  A.~V.~Sologubenko, {\it et al}.: 
Phys. Rev. B {\bf 62} (2000) R6108.
%
\bibitem{Sol2} A.~V.~Sologubenko, {\it et al}.: 
Phys. Rev. B. {\bf 84} (2001) 054412.
%
\bibitem{SODR} A.~V.~Sologubenko, {\it et al.}:
Europhys. Lett. {\bf 62} (2003) 540.
%
%
\bibitem{ZP03} For a review see X.~Zotos and P.~Prelov\v sek:
  cond-mat/0304630.
%
\bibitem{znp} X.~Zotos, F.~Naef and P.~Prelov\v sek: Phys. Rev. 
B {\bf 55} (1997) 11029.
%
\bibitem{KS} A.~Kl\"umper and K.~Sakai:
J. Phys. A {\bf 35} (2002) 2173.
%
\bibitem{SK}  K.~Sakai and  A.~Kl\"umper:
J. Phys. A {\bf 36} (2003) 11617.
%
%
\bibitem{NarMA98} B.~N.~Narozhny, A.~J.~Millis and N.~Andrei: Phys. Rev. B
{\bf 58}  (1998) R2921.
%
\bibitem{Zotos99} X.~Zotos: Phys. Rev. Lett. {\bf 82} (1999) 1764.
%
%
\bibitem{AlGros1} J.~V.~Alvarez and C.~Gros: Phys. Rev. Lett. {\bf 88} (2002)
077203.
%
\bibitem{FK03} S.~Fujimoto and N.~Kawakami: Phys. Rev. Lett. {\bf 90} (2003)
 197202.
%
%
\bibitem{Kudo} K.~Kudo, {\it et al.}: 
J. Low. Temp. Phys. {\bf 117} (1999) 1689.
%
\bibitem{Sol3} A.~V.~Sologubenko, {\it et al}.: 
Phys. Rev. Lett. {\bf 84} (2000) 2714.
%
\bibitem{Hess01} C.~Hess,{\it et al.}: Phys. Rev. 
B {\bf 64} (2001) 184305.
%
\bibitem{Kudo01} K.~Kudo, {\it et al.}: 
J. Phys. Soc. Jpn. {\bf 70} (2001) 437.
%
\bibitem{MHCB} F.~Heidrich-Meisner,{\it et al.}:
Phys. Rev. B {\bf 66}, 140406(R) (2002).
%
%
\bibitem{AlGros2} J.~V.~Alvarez and C.~Gros, Phys. Rev. Lett. {\bf 89},
156603 (2002).
%
\bibitem{OCC03} E.~Orignac, R. Chitra and R.Citro, Phys. Rev. B
{\bf 67}, 134426 (2003).
%
\bibitem{Saito1} K.~Saito and S.~Miyashita, J. Phys. Soc. Jpn. {\bf 71},
2485 (2002).
%
\bibitem{Saito2} K.~Saito, Phys. Rev. B {\bf 67}, 064410 (2003).
%
\bibitem{Saito3} K.~Saito, Phys. Rev. B {\bf 67}, 164410 (2003).
%
%
\bibitem{MHCB2}  F.~Heidrich-Meisner, A.~Honecker,
D.C.~Cabra and W.~Brenig, Phys. Rev. B {\bf 68} (2003) 134436.
%
\bibitem{SAR03} E.~Shimshoni, N.~Andrei and A.~Rosch,
Phys. Rev. B {\bf 68} 104401 (2003).
%
\bibitem{Louis03} K.~Louis and C.~Gros, Phys. Rev. B {\bf 68}
(2003) 184424.
%
\bibitem{Zotos04} X.~Zotos, Phys. Rev. Lett. {\bf 92} (2004) 067202.
%
\bibitem{KZ04} J.~Karadamoglou and X.~Zotos, Phys. Rev. Lett.
{\bf 93} (2004) 177203.
%
\bibitem{LG03} K.~Louis and C.~Gros: Phys. Rev. B {\bf 67} (2003) 224410.
%
\bibitem{MHCB3}  F.~Heidrich-Meisner, {\it et al.}: 
Phys. Rev. B {\bf 71}, 184415 (2005).
%
\bibitem{SK04} K.~Sakai, A.~Kl\"umper: J. Phys. Soc. Jpn. {\bf 74} 
(2005) Suppl. 196. cond-mat/0410192.
%
\bibitem{Takabook} M.~Takahashi, {\it Thermodynamics of One-Dimensional
Solvable Models}, (Cambridge: Cambridge University Press).
%
\bibitem{FK98} S.~Fujimoto and N.~Kawakami, J. Phys. A {\bf 31}, 465 (1998).
%
\bibitem{EKSstring}  H.L.~E{\ss}ler, V.E.~Korepin and K.~Schoutens,
              J. Phys. A {\bf 25} (1992) 4115.
%
\bibitem{JuDo} G.~J{\"u}ttner and B.D. D{\"o}rfel,
           J. Phys. A {\bf 26} (1993) 3105. 

\bibitem{AlczMart}  F.C.~Alcaraz and M.J.~Martins, J. Phys. A{\bf 21} 
                  (1988) L381,
                   ibid. 4397.

%

%
%
\bibitem{CB} P.~M.~Chaikin, G.~Beni, Phys. Rev. B {\bf13} (1976) 647.
%
\bibitem{MP04} M.~M.~Zemljic and P.~Prelov\v sek:
 Phys. Rev. B {\bf 71} (2005) 085110.
\end{thebibliography}
\end{document}